# A photon calorimeter using lead tungstate crystals for the CEBAF Hall A Compton polarimeter

D. Neyret[1], T. Pussieux, T. Auger, M. Baylac, E. Burtin, C. Cavata,
R. Chipaux, S. Escoffier, N. Falletto, J. Jardillier, S. Kerhoas,
D. Lhuillier, F. Marie, C. Veyssière

*CEA Saclay, DAPNIA*
*91191 Gif sur Yvette Cedex, France*

J. Ahrens, R. Beck, M. Lang

*Institut für Kernphysik*
*Johannes Gutenberg-Universität Mainz*
*D-55099 Mainz, Germany*

A new Compton polarimeter is built on the CEBAF Hall A electron beam line. Performances of 10% resolution and 1% calibration are required for the photon calorimeter of this polarimeter. This calorimeter is built with lead tungstate scintillators coming from the CMS electromagnetic calorimeter R&D. Beam tests of this detector have been made using the tagged photon beam line at MAMI, Mainz, and a resolution of $1.76\% \oplus \frac{2.75\%}{\sqrt{E}} \oplus \frac{0.41\%}{E}$ has been measured.

---

1. Corresponding author. Tel: (33-1) 69 08 75 52; fax: (33-1) 69 08 75 84; e-mail: dneyret@cea.fr

# 1.0 Introduction

CEBAF[1] is an electron accelerator located at Jefferson Laboratory (Newport News, Virginia, USA). It delivers continuous electron beam from 0.8 to 6 GeV to three experimental halls with a current up to 100 µA. A polarized source can be used to produce a beam polarized up to 70%. Some experiments and in particular the HAPPEX parity violation experiment [1] require a fast (less than an hour) and accurate (3%) measurement of the beam polarization. Two polarimeters are already used for this measurement: a Mott polarimeter located at the injector, and a Møller polarimeter on the Hall A beam line (limited to low current, 10 µA). But both techniques are destructive for the beam properties and can not be operated simultaneously with the physics experiments.

Saclay´s group, in collaboration with the Clermont-Ferrand LPC and the Jefferson Laboratory, is building on the Hall A beam line a Compton polarimeter which will give a non destructive measurement for up to 100 µA beam currents. This is done by shooting an IR laser beam on the electron beam, detecting and measuring the energy of the photons scattered after the Compton interaction. The counting rate asymmetry of this interaction between two beam polarization states depends on its kinematics and on the beam polarization.

We will first give a brief description of the Compton polarimetry, the performances required for the instrument and more specifically for the photon calorimeter. We will then present the tests performed at the Mainz tagged photons beam and the results on the lead tungstate crystals resolution at the energy level used with the polarimeter.

# 2.0 Compton polarimetry

## 2.1 Polarization measurement

In the Compton polarimeter, a circularly polarized infra-red laser beam interacts with the polarized electron beam via the Compton process, and an asymmetry is measured when the helicity state of the electron beam changes. The electron beam polarization can be extracted from this asymmetry, where $n_+$ and $n_-$ are the Compton counting rates for the two polarization states:

$$A_{\exp} = \frac{n_+ - n_-}{n_+ + n_-} = P_e P_\gamma A_l$$

The laser beam polarization $P_\gamma$ is measured and the theoretical asymmetry $A_l$ is calculated with QED and depends on the scattered photon energy. This asymmetry is calculated from the Compton cross section with spin parallel and anti-parallel:

$$A_l = \frac{\sigma_{\Rightarrow}^{\rightarrow} - \sigma_{\Rightarrow}^{\leftarrow}}{\sigma_{\Rightarrow}^{\rightarrow} + \sigma_{\Rightarrow}^{\leftarrow}}$$

If $\rho = k / k_{max}$ is a kinematics parameter, where $k$ and $k_{max}$ are respectively the current and maximum energy of the scattered photon, the differential unpolarized cross section, with a crossing angle equal to zero, is [3] [4]:

$$\frac{d\sigma}{d\rho} = 2\pi r_0^2 a \left[ \frac{\rho^2 (1-a)^2}{1 - \rho(1-a)} + 1 + \left(\frac{1 - \rho(1+a)}{1 - \rho(1-a)}\right)^2 \right]$$

---

1. Continuous Electron Beam Accelerator Facility

where $r_0$ is the classical electron radius, and $a = 1 / (1+4kE/m^2)$ with k and E the energies of the initial photon and electron, and m the electron mass (Figure 1). The total cross section for all the $\rho$ range with 4 GeV electrons and 1.16 eV IR laser photons reach a value of 0.62 barn.

The theoretical longitudinal differential asymmetry is given by:

$$A_l = \frac{2\pi r_0^2 a}{\frac{d\sigma}{d\rho}}(1 - \rho(1 + a))\left[1 - \frac{1}{(1 - \rho(1 - a))^2}\right]$$

This asymmetry is shown in Figure 2. For 4 GeV electron beam and 1.16 eV photons, the energy of the scattered photons ranges between 0 to 250 MeV. We can see that at CEBAF energy level, the theoretical asymmetry is small (less than 8% at 4 GeV electron beam). This implies that a large amount of events is required to get a good statistical accuracy. One can also note that this asymmetry is negative at low scattered photon energy and positive at high energy. Even if the integrated asymmetry is non-vanishing, a better accuracy is reached for the same amount of events, by measuring the energy for each Compton event. Indeed, for N acquired events, the statistical error for the electron polarization with an integrated method (where the $\rho$ value is not measured) is [2]:

$$\left(\frac{\Delta P_e}{P_e}\right)^2 \approx \frac{1}{N P_e^2 P_\gamma^2 \langle A_l \rangle^2}$$

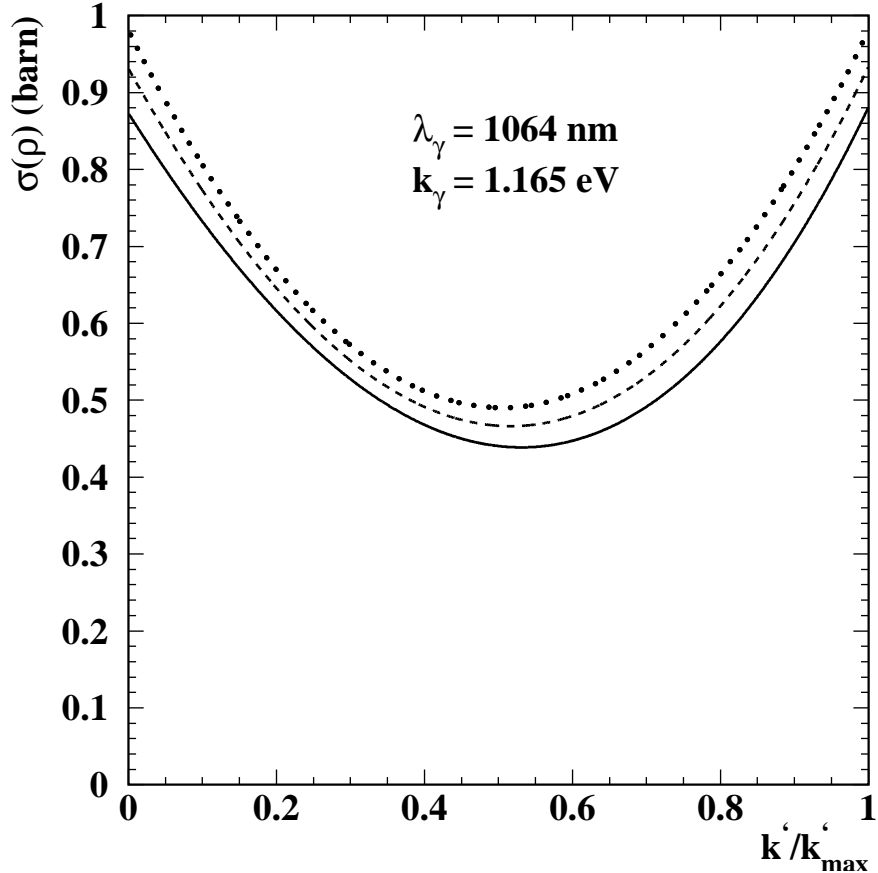

**FIGURE 1.** Compton differential unpolarized cross section according to the scattered photon energy for 1 GeV (dotted, $k'_{max}$ = 18 MeV), 4 GeV (dashed, $k'_{max}$ = 266 MeV) and 8 GeV (solid line, $k'_{max}$ = 999 MeV) electron beam

If the ρ parameter is measured for each event, a differential measurement could be used, and the statistical error in this case is given by:

$$\left(\frac{\Delta P_e}{P_e}\right)^2 \approx \frac{1}{N P_e^2 P_\gamma^2 \langle A_l^2 \rangle}$$

In both cases, the average is calculated on the kinematics acceptance region. For the realistic detection domain (ρ between 0.2 and 1) the number of events needed to reach a statistical error level is 3 times lower for the latter method. For example with a 4 GeV electron beam with 50% polarization, the differential method requires $35.10^6$ events to reach 1% statistical error ($\sqrt{\langle A_l^2 \rangle}$ near 3.8%), whereas the integrated method requires $102.10^6$ events ($<A_l>$ near 2.3%). However, this amount of events is quite large. It is therefore mandatory to have a high counting rate (up to 100 kHz) to perform a polarization measurement in within half an hour.

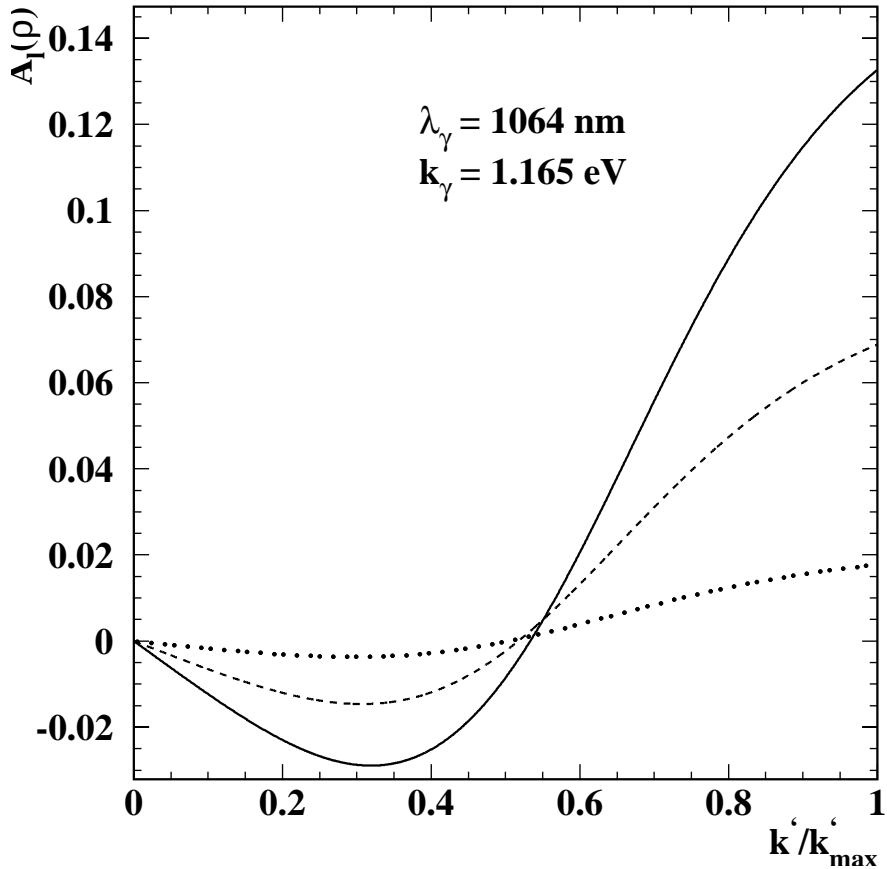

**FIGURE 2.** Theoretical asymmetry for CEBAF energies (1 GeV dotted, 4 GeV dashed, 8 GeV solid line) according to the scattered photon energy

In practice, for the differential method we divide the energy range into N bins $[\rho_i, \rho_{i+1}]$. The electron polarization is calculated in each bin with:

$$P^i_{ex} = \frac{A^i_{ex}}{A^i_{th} P_\gamma}$$

$$\text{where} \quad A^i_{ex} = \frac{N^i_+ - N^i_-}{N^i_+ + N^i_-} \quad \text{and} \quad A^i_{th} = \frac{\int_{\rho_i}^{\rho_{i+1}} A_l \frac{d\sigma}{d\rho} d\rho}{\int_{\rho_i}^{\rho_{i+1}} \frac{d\sigma}{d\rho} d\rho}$$

The final electron polarization is then the weighted mean of the bin polarizations:

$$P^{DM}_{ex} = \frac{\sum_{i=1}^{N} \frac{P^i_{ex}}{(dP^i_{ex})^2}}{\sum_{i=1}^{N} \frac{1}{(dP^i_{ex})^2}}$$

where $dP^i_{ex}$ is the statistical accuracy of the polarization measurement in each bin:

$$dP_i = \frac{2\sqrt{N^+_i N^-_i (N^+_i + N^-_i)}}{(N^+_i + N^-_i)^2}$$

## 2.2 Effects of the detector resolution and calibration

When using the differential method, the $\rho$ value has to be measured for each event using the photon calorimeter. Measurement errors (resolution and miscalibration) have some effects on the measured asymmetries and can affect the polarization measurement.

Resolution effects $\sigma_{res}(\rho)$ can smear the cross section slope which can behave like (for example with a gaussian law):

$$\frac{d\sigma_s}{d\rho_s}(\rho_s) = \int_0^1 \frac{d\sigma}{d\rho}(\rho) \frac{1}{\sqrt{2\pi}\sigma_{res}(\rho)} e^{-\frac{(\rho - \rho_s)^2}{2\sigma_{res}(\rho)^2}} d\rho$$

The smearing effect then change the value of $A^i_{th}$ for each bin which becomes:

$$A^i_{sm} = \frac{\int_{\rho_i}^{\rho_{i+1}} A_l \frac{d\sigma_s}{d\rho_s} d\rho_s}{\int_{\rho_i}^{\rho_{i+1}} \frac{d\sigma_s}{d\rho_s} d\rho_s}$$

With the exact calculation of $A^i_{sm}$, the electron polarization can be measured. The resolution of the measurement only affects the amount of data needed to have a good statistical error. With 10% resolution on average, this amount is just 10% higher than with a precise measurement: a 10% resolution therefore fulfills our requirements. The tricky part is to estimate accurately the detector resolution in order to calculate $A^i_{sm}$ with a good precision. This value can be calculated if the detector resolution is known on all the energy range. A poor knowledge of this resolution can lead to an error on the polarization measurement. Some studies [2] have been done on this effect, using a detector resolution value of $\frac{\sigma(E)}{E} = 1\% \oplus \frac{3\%}{\sqrt{E}} \oplus \frac{1\%}{E}$, with 50% for the one photoelectron resolution. These studies have shown that the most sensible term is the statistical one; an uncertainty of 10% on this parameter lead to an error on the electron polarization of 1%.

Another error arises from the miscalibration of the detector. This effect has been studied by using a linear relation between the measured value $\rho_m$ and $\rho_s$, with $\rho_m = \rho_0 + (1+s)\rho_s$ where $\rho_0$ represents a systematic shift and s a slope miscalibration. These studies have shown that an error of 1% on the slope (s=0.01) lead to a polarization measurement error of 1%, and a systematic shift of $\rho_0$=0.01 leads to an error of 2% on the polarization.

Eventually, the electron polarization measurement can be done with an average resolution of 10% only, but we need to know this resolution with a 10% accuracy. Moreover, the detector has to be calibrated with a precision of 0.5% to have a beam polarization measurement at 1% precision level.

### 3.0 The Hall A Compton polarimeter

In the Compton process at our kinematics, both particles are scattered at small angles compare to the electron initial direction (less than 500 µrad for photons and less than 5 µrad for electrons). To separate the scattered photons from the electron beam, we use a magnetic chicane (see Figure 3) built with four magnetic dipoles [2]. The laser beam is set between the second and the third dipole. The photon calorimeter is set in the line of sight of this segment, just before the fourth dipole. An electron detector will be installed after the third dipole and will measure the deviation of the scattered electron from the electron beam. Figure 4 gives a schematic view of the polarimeter.

We have seen that we need a high luminosity to have a good statistical accuracy. To produce Compton events with a high counting rate, the laser beam must deliver a high power continuously (more than 500 W). Such lasers are very expensive and would be very difficult to install in the experimental area. So, in order to reach such a high power, we amplify a low power infra-red laser (300 mW) with a Fabry-Pérot cavity [5]. This cavity amplifies the laser beam with a gain of 3000, which gives an equivalent power of 900 W. With this setup, the counting rate of Compton events reaches more than 100 kHz with nominal electron beam conditions.

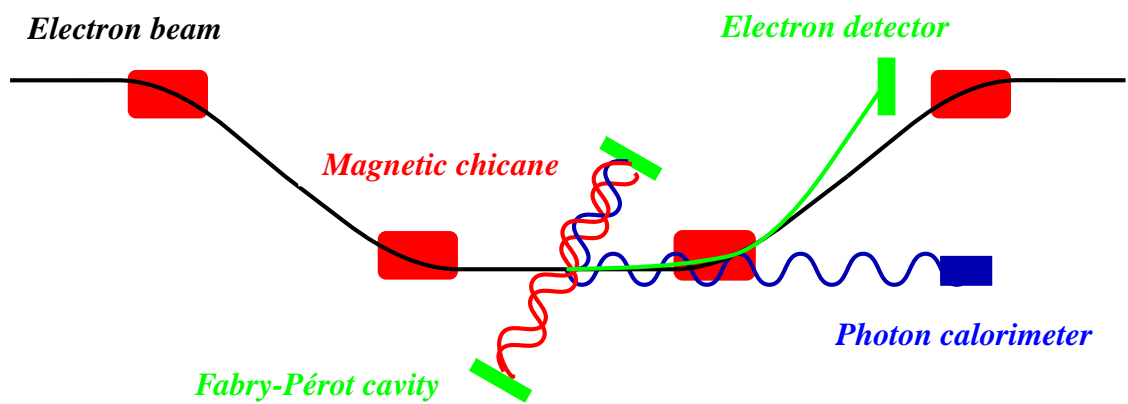

**FIGURE 3.** Sketch of the CEBAF Hall A Compton polarimeter

## 4.0 The photon calorimeter

The Compton scattering produces a beam of scattered photons with an energy between 0 to 250 MeV for 4 GeV electron beam at a counting rate of 100 kHz (with 100 µA 4 GeV electron beam and 500 W laser beam). We must measure the energy of each of those photons, at this counting rate, with a calibration of 0.5% and an average resolution of 10%. The space available for the detector is very small, as the distance between the scattered photon beam and the electron beam is only 30 cm near the fourth dipole. The environment is very radiative (synchrotron radioation), the detector must therefore be radiation hard.

A good candidate to build this calorimeter is the $PbWO_4$ scintillator [6] which is issued from the R&D for the CMS electromagnetic calorimeter [7] [8]. It is dense (8.28 g/cm$^3$) with a small Molière radius (2.19 cm), so we could build a small detector. It is fast (85% of the charge in 25 ns), the detector occupation can be very short (some 100 ns) reducing the dead time at high counting rate and the pile-up effects. Its light yield is greater than 6 photoelectrons (γe) by MeV of deposited energy, which seems enough to reach a 10% resolution even at low energy. Finally, it is not very sensitive to radiation [9], so its characteristics don't change with time.

The final photon calorimeter set-up is shown on Figure 5. It is composed of a matrix of 5 by 5 lead tungstate crystals (size 2 x 2 x 23 cm) doped with niobium coming from Bogoroditsk plant [10]. These crystals are wrapped with Tyvek paper to isolate them optically and they are read by 25 photomultipliers tubes[1] supplied with positive high voltage. They are monitored with a blue diode[2] which sends fast flashes (5 ns) in the crystals via optical fibers. As the crystals light yield can vary with the temperature of 2%/ºC, they are thermalized at 16ºC and temperature probes monitor the crystals temperature.

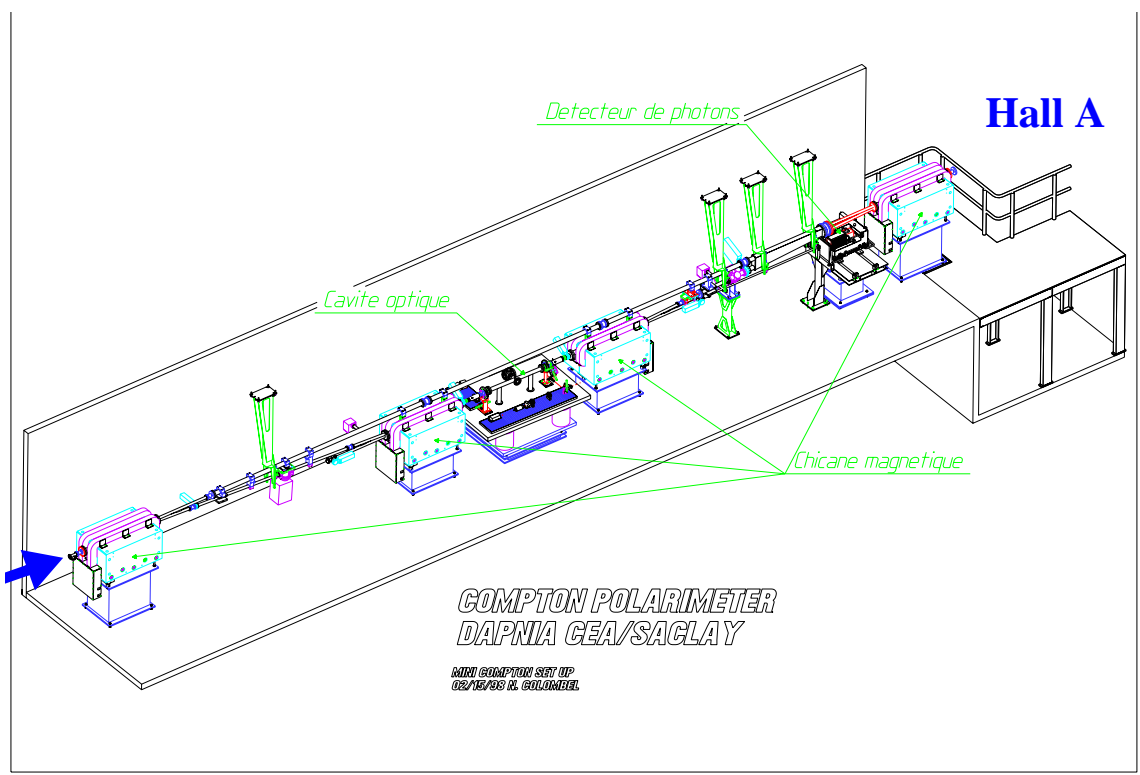

**FIGURE 4.**     Plan of the Compton polarimeter

---

1. Philips XP1911 19 mm diameter 85 mm long
2. Ledtronics 525nm Ultra aqua green BP280CWAG6K-3.5V-050l

## 5.0 Tests at Mainz with low energy photons

Tests were done in order to check that these crystals fulfill the polarimeter requirements. In 1997, we received 25 crystals from Bogoroditsk plant. Light yield and decay time measurements were performed at Saclay [11]. The decay time measurements showed that in addition of the standard fast decays around 5, 20 and 100 ns, there seems to be a slow component around 1 µs. However this component is difficult to measure and the pile-up effect it can cause is negligible at 100 kHz counting rate. The light yield measurements showed a light yield between 6 to 8 γe/MeV.

Beam tests have been done on the whole detector with 5 by 5 crystal matrix. They were performed on the tagged photon beam at Mainz [12]. The 855 MeV MAMI electron beam is sent to a radiator target. The scattered electrons are deviated by a spectrometer dipole and detected by scintillators plates. With this setup, the energy of each bremsstrahlung photon is known with a 2 MeV accuracy. The spectrometer can tag scattered photons between 40 to 800 MeV.

The goal of these tests was to determine the characteristics of the detector at the Compton energies (between 40 to 400 MeV), and in particular to measure the resolution of the calorimeter as a function of the photon energy.

The first step was to calibrate individually each crystal, in relation to the signal given by the monitoring device to compensate the PMT variations. For this, each crystal (i) has been placed in front of the beam and its response $Q_i$ has been compared to the photon energy $E_\gamma$. Figure 6 shows the ADC spectrum of the crystal 13 for the monitoring diode and for different photon energies. These functions have been modelled by a straight line and divided by the monitoring diode responses $Q^d_i$:

$$Q_i^{\text{norm}} = \frac{Q_i - P_i}{Q_i^d - P_i} = \alpha_i E_{\text{left}_i} = \alpha_i \lambda E_\gamma$$

with $E_{\text{left}}$ of the energy left in the crystal by the photon, $\lambda$ the ratio of this energy compared to the initial energy (we assume $\lambda$ to be crystal independent), $ped_i$ the pedestal of the electronic system and $\alpha_i$ the linear factor. This scan thus gives the values of $\alpha_i \lambda$.

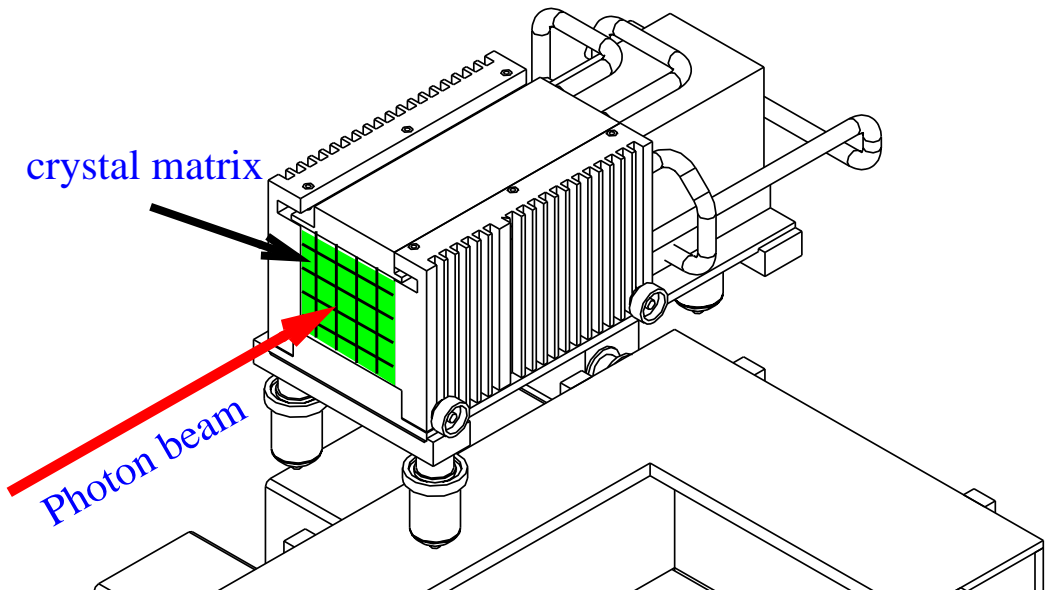

**FIGURE 5.**     Photon calorimeter set-up

During these tests, parasitic electronic noises appeared between the PMT and the electronic front-end with a frequency near 40 kHz. The effects of those noises were to smear the pedestals (with a sigma near 3.5 ADC channels, that is to say 1.8 MeV) and signal spectrums. These noises were almost totally correlated from one channel to the other. To cancel them to the maximum, we have used a PMT which was not linked to any crystal. The correlations of the noise between this control PMT and the others have been studied with random events used for pedestal calculations. With these correlations, the ADC values of all the PMT for each event have been corrected with the ADC value of the control PMT. For the future data taking at CEBAF, cable shielding have been greatly improved to reduce this noise by a factor 10.

The central crystal has eventually been put directly into the photon beam to study the response of the whole detector. For this position, the responses of the crystals have been normalized using the precedent scan and the diode response, and then added:

$$E_{\mathrm{mes}} = \sum E_{\mathrm{left}_i} = \sum \frac{Q_i - P_i}{\alpha_i (Q_i^d - P_i)}$$

Figure 7 shows the responses ($E_{\mathrm{mes}}/\lambda$) of the central crystal, of the sum of the 9 crystals in the center of the matrix, and of the whole matrix. The linearity versus the gamma energy is better than 1%. If we assume that the entire photon energy is deposited in the crystal matrix, the $E_{\mathrm{mes}}$ value is equal to $E_\gamma$ and the value of $\lambda$ can be measured. With this measurement, $\lambda$ is equal to 72%, which means that 72% of the total deposited energy is found in the central crystal. Moreover 22% is found in the 8 crystals of the inner crown, and 6% in the rest.

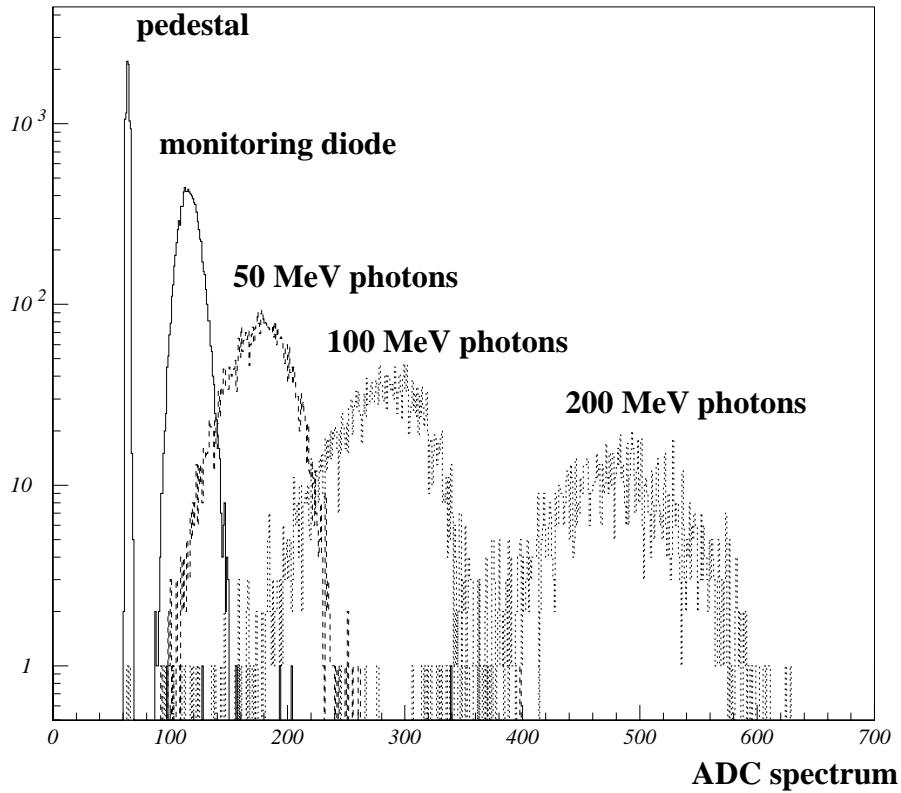

**FIGURE 6.** $V_{\mathrm{ADC}}$ spectrum for the central crystal for random events (pedestal), for monitoring diode, for 50 MeV, 100 MeV and 200 MeV photons

The resolution (defined as $\sigma(E_{mes})/<E_{mes}>$ where $\sigma$ is the sigma of the $E_{mes}$ distribution) versus the photon energy is shown on Figure 8, for both 9 crystals and 25 crystals configurations. The resolution for 25 crystals is lower than 10% for gamma energy greater than 100 MeV, and lower than 20% for gamma above 40 MeV, which is sufficient for our needs. These resolution distributions have been fitted by the classical function $\frac{\sigma}{E} = a \oplus \frac{b}{\sqrt{E}} \oplus \frac{c}{E}$, where a is associated with noise proportional to the energy (calibration error, physics noise), b is associated with the light output of the crystals and c with electronic noises independent of the photon energy.

The parameters of the fits are given in Table 1. The b value for the whole detector corresponds to a light output of 1.3 γe/MeV, and for 9 crystals, it corresponds to 1 γe/MeV. The value given for central crystal is not significant due to the bad quality of the fit.

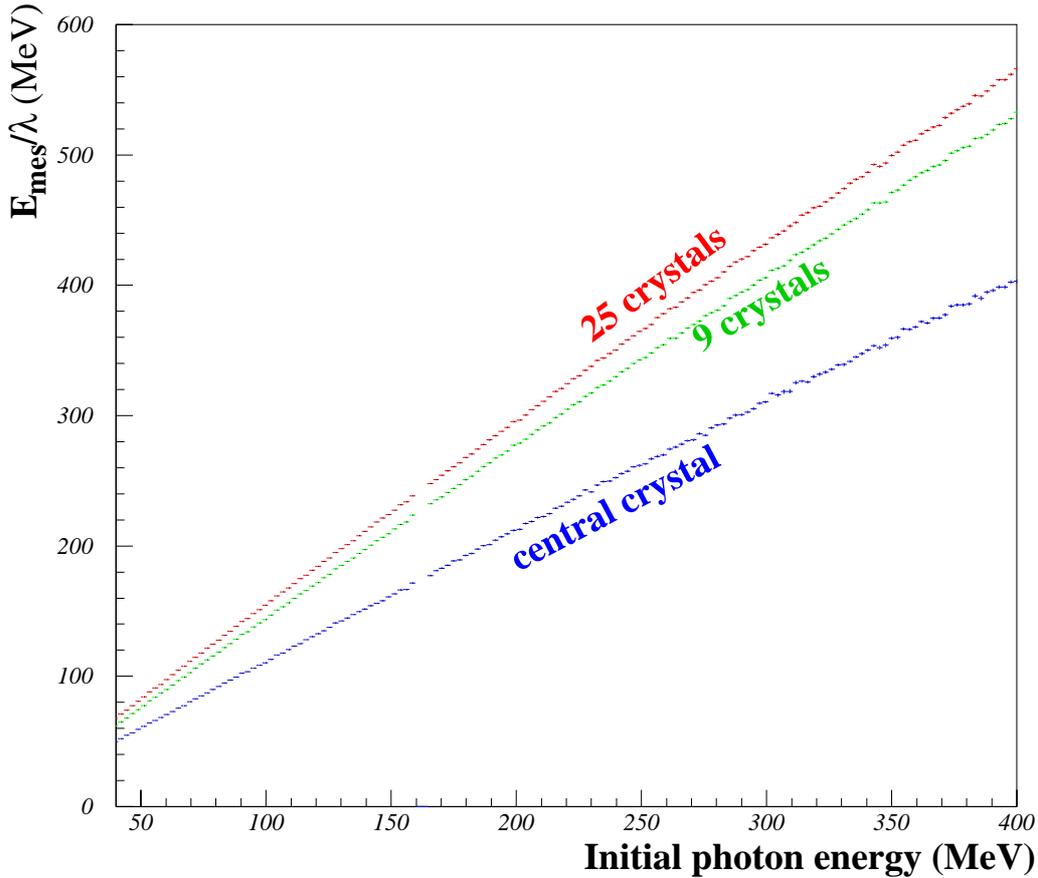

**FIGURE 7.** Response $\left( \sum \frac{Q_i - P_i}{\alpha_i \lambda (Q_i^d - P_i)} = \frac{E_{mes}}{\lambda} \right)$ of the calorimeter for tagged photons between 40 to 400 MeV with all the crystals, the 9 central crystals and the central crystal only

## Conclusion

The beam test made on the photon calorimeter at the MAMI tagged photon beam show that its characteristics meet our requirements, with a resolution of the entire calorimeter (25 crystals) of $1.76\% \oplus \frac{2.75\%}{\sqrt{E}} \oplus \frac{0.41\%}{E}$. It can allow a polarization measurement of the electron beam with a good accuracy within a small measurement period. A large part of the polarimeter has been installed and commissioned in 1998, except for the optical cavity which will amplify the laser beam. Due to the low power of the laser used in 1998, the Compton events counting rate was too low compared to the background to proceed to the polarization measurement. Some studies are performed to reduce this background and, at the beginning of 1999, the optical cavity will be installed. This will allow a Compton counting rate much greater than the background rate. A new data acquisition system will also be installed in February 1999, this system will acquire up to 100 kHz events. We expect to measure the beam polarization continuously at 3% total precision during the next HAPPEX Parity violation experiment [13] between April and June 1999.

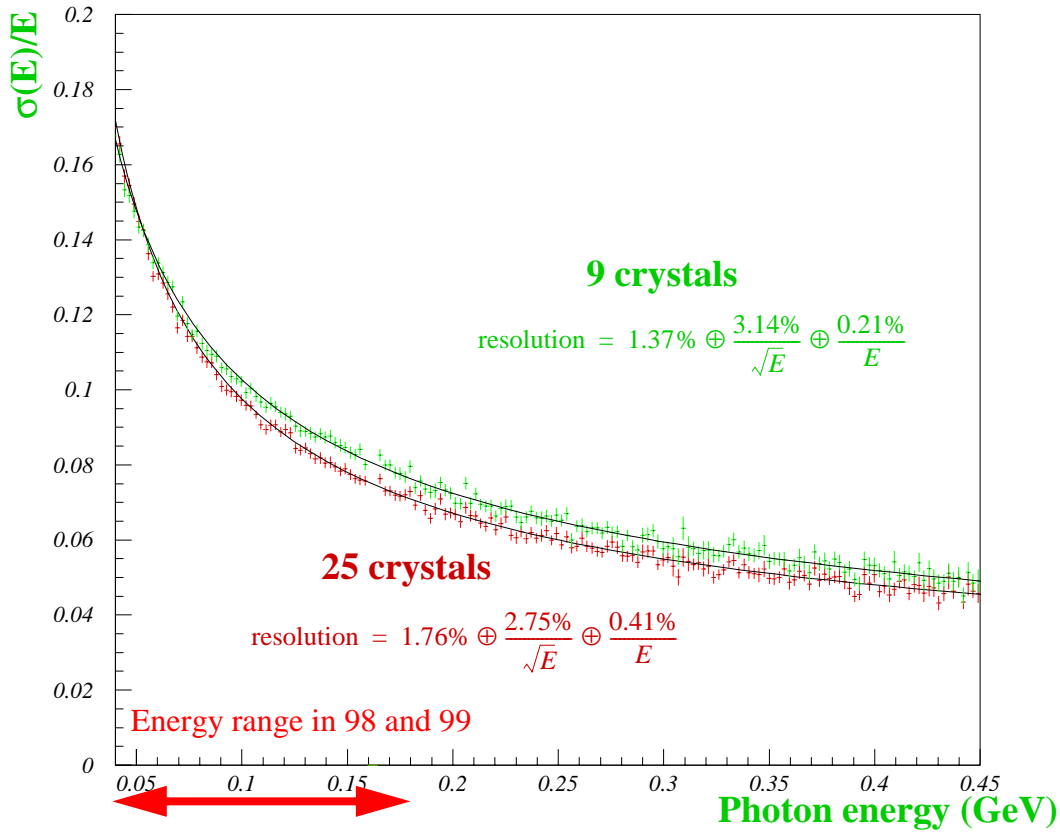

**FIGURE 8.** Resolution of the detector for 9 and 25 crystals taken in account according the initial photon energy

|                 | a(%)          | b(%)         | c(%)           | $\chi^2$/ndf |
|-----------------|---------------|--------------|----------------|--------------|
| central crystal | 4.48 ± 0.11   | 4.80 ± 0.02  | 0 ± 0.2        | 3.67 / 157   |
| 9 crystals      | 1.37 ± 0.25   | 3.14 ± 0.03  | 0.21 ± 0.03    | 0.77 / 167   |
| 25 crystals     | 1.76 ± 0.17   | 2.75 ± 0.03  | 0.41 ± 0.014   | 0.77 / 167   |

**TABLE 1.** Parameters of the resolution slope with fit errors, $\chi^2$ and number of degrees of freedom of the fit for central, 9 and 25 crystals, with

$$\text{resolution} = a \oplus \frac{b}{\sqrt{E}} \oplus \frac{c}{E},$$ where E is the initial photon energy